\documentstyle[11pt,psfig]{article}
\newcommand{\fig}[1]{Fig.~{#1}}
\newcommand{\eqn}[1]{Eq.~({#1})}
\newcommand{\secn}[1]{Sec.~{#1}}
\newcommand{\subn}[1]{Sec.~{#1}}

\title{Dynamics of a Grafted Chain into a Rubber Network:
	A Monte Carlo Study} 
\author{J.M. Deutsch \quad Hyoungsoo Yoon \\
	University of California, Santa Cruz \\
	Santa Cruz, CA 95064}
\date{April 8th 1994\\
	cond-mat/9404012}

\begin{document}
\maketitle

\begin{abstract}
The dynamics of a single chain tethered to an interface
and in contact with a cross-linked network is examined numerically.
When the network is put in contact with the tethered chain, the chain
moves with dynamics that are highly constrained
due to entanglements. When the surface is repulsive, the chain
runs straight along the surface and then forms 
a plume in the network that starts at a
distance of order $\sim\!N^{1/2}$ from the graft point.
For short times, the chain length in the gel
increases algebraically as a function of
time, in most cases as $\sim\!t^{1/2}$.
The plume configuration is highly metastable and on a much
longer time scale the point of entry into the network decreases
to zero. This is similar to the relaxation of the arm of a star polymer
in a cross-linked network. The above findings are in agreement
with the analytical predictions of O'Connor and McLeish. 
The effects of a chemical disparity between
the grafted chain, the network, and the substrate are investigated.
Topological constraints are placed in the interface to determine
their effects on the dynamics.
Chains tethered at both ends are also studied and show a transition
in behavior as a function of the thickness of the interface.
Above a critical thickness chain does not penetrate.

\end{abstract}

\section{Introduction}
\label{introduction}

\subsection{Adhesion via connectors}
\label{adhesion_via_connectors}

Polymer chains grafted to a surface or an interface have been the
focus of extensive study
during the past 15 years because of their practical
importance.
In particular, polymers grafted at the interface of two polymeric
networks or solids, often diffuse into the
bulk which will then increase the bonding of the two solids.
This may
be due to some type of chemical bonding
between the grafted chains and the bulk polymers
but can also be due to
entanglements between them.
Grafted polymers with such functions have been termed 
{\em connectors}
for this reason\cite{ERPGdG:RubberRubber}.

Since the junction is weaker than the bulk, the interface region
usually fractures when an excessive stress is 
applied\cite{PGdG:WeakAdhesive}.
As the crack tip propagates along the interface, the chains 
either break or are
pulled out, and this process is believed to
contribute to the toughness of
the interface\cite{HRB:AdhesionPolymers}.

Although there have been numerous studies of the molecular
mechanism of fracture for these systems,
relatively little is known
about the dynamics of the interface's formation.
Understanding the dynamical aspects of this problem is
important. As will be discussed below, the approach to equilibrium
of these systems are often so slow that equilibrium
conformations
can never be achieved in the time scale of the experiments.
Therefore the dynamics of these systems must be considered.
Different models of chain pull-out, which make predictions about
the toughness of an interface, make different assumptions
about the dynamics. Since there are some discrepancies between 
different 
models\cite{ERPGdG:RubberRubber,PGdG:WeakAdhesive,%
HJPGdG:AdhesionConnector,KEE:ScalingAnalysis}
it is important to have a better theoretical understanding of
the dynamics, as well as performing further experiments.
Another important application of the interface dynamics
is in understanding the case of interface ``healing''.
In the case where the toughness arises solely from 
entanglement effects,
an interface can be healed even after a complete
physical failure. To understand the recovery of  the toughness
upon re-contact requires a detailed knowledge of the system's
dynamics.

Recently O'Connor and McLeish\cite{KPOTCBM:MolecularVelcro},
motivated by the experiment of Reichert
and Brown\cite{WFRHRB:ReicherAndBrown},
investigated the relaxation of connector chains in an
elastomer network attached to a glassy polymer.
They considered the case where
the chains are compatible with the elastomer
but not with the surface of the interface.
They found that there exist three stages in healing upon
contact with the interface.
First, the chain relaxes inside the interface. 
After this, the chain penetrates into the network.
This is achieved in the Rouse time and most of the toughness is
recovered at this stage.
On a much longer time scale,
the penetration point moves logarithmically toward the
point where the chain is anchored.
The typical value of the total toughness gain by this
{\em penetration-point-hopping\/} 
process was estimated to be about 2.

Soon after the theoretical work of O'Connor and McLeish,
Creton et~al.\cite{CCHRBetal:MolecularWeight}
conducted a more detailed experiment with some of the
results apparently inconsistent with the current theories 
(See also Brown\cite{HRB:EffectsChain}).
They carried out an experiment,
which was originally devised by Johnson, Kendall and 
Roberts\cite{KLJKKetal:JKRExperiment},
on the interface between a cross-linked polyisoprene~(PI)
matrix and a polystyrene~(PS) substrate grafted
with PI chains.
One interesting result of their work is that 
the dependence of toughness on molecular weight
of tethered chains is not as strong as theories predict.
It is hoped 
that a more complete understanding of the dynamics of the grafted
chains will reconcile some of these inconsistencies.

\subsection{Chain pull-out Models}
\label{chain_pull-out_models}

In this paper we consider the interface formed between a cross-linked
polymer matrix and a rigid, non-reactive surface when
connector chains, with polymerization index~$N$, 
are end-tethered to the surface.
The chains are assumed to be chemically compatible with the matrix
unless specified otherwise and have a monomer size or {\em Kuhn length}
of~$a$.

The degree of the coverage is usually classified, according to the
surface density, $\sigma$, of the grafted chains.
At low densities, $\sigma \ll 1/a^2N$,
one has what is termed a ``mushroom regime" since single
chains that are repelled by a surface form a mushroom-like
or plume-like shape in equilibrium. 

In this paper we study 
this low density regime where the interaction between
different connector chains can be ignored, in which case we
only need to consider the  behavior of a single connector chain.
The aim of the present work is to study the relaxation of a  
connector chain from a conformation lying on the surface 
to the equilibrium plume-like configuration. 
Experimentally 
relevant quantities are obtained by averaging over the
ensemble of these (non-interacting) chains.

What is the most relevant quantity one needs to measure in order
to describe the toughness enhancement of the interface 
arising from the pull-out of a single connector chain?
First we note that the chain pull-out process is viscous. 
Hence toughness enhancement due to the connectors
should be a monotonically increasing function of the chain length
inside the rubber.
When the crack propagation speed is larger than a certain critical value,
a simple theory\cite{KEE:ScalingAnalysis}
predicts that the toughness due to this 
energy loss when connector chain is
pulled out, is proportional to the square of the chain length
inside the rubber.
When this speed is small Rapha\"{e}l and
de~Gennes\cite{ERPGdG:RubberRubber,%
HRBCYHetal:InterplayIntermolecular}
formulated a theory which predicts
that the toughness is linearly proportional to the chain length
inside the rubber.
As such, the chain length
inside the rubber
is expected
to play an important role in describing the toughness recovery
process,
and is therefore
measured as a function of time in our simulation.

The rest of the paper is organized as follows.
In the next section, we will give the precise definition of the problem
under investigation and
some important concepts will be introduced.
After this the numerical model and the simulation
algorithm will be given in detail.

Our simulation results will be discussed
in the following sections.
The initial expansion process after the contact 
of the two materials, is studied in~%
\secn{\ref{chains_tethered_at_both_ends}}.
We study the dynamics of the chain with both ends tethered as
well as that of the chain with only one end fixed. 
The double grafting is
interesting not only because of its relevance to the initial
relaxation of the singly grafted chain, but also because this might
have some interesting experimental applications.
\secn{\ref{chains_tethered_at_one_end}} 
considers the relaxation of the free end into the rubber.
First, the theoretical model of O'Connor and McLeish will be 
discussed in~\subn{\ref{theory}}.
The effects of chemical disparity between the chain
and the matrix or between the chain and the substrate
is also discussed here.
Simulation results are presented in~\subn{\ref{simulation_results}}.
\secn{\ref{interfacial entanglement_effects}} 
studies a situation different from 
the ones above in that the
dynamics of the grafted chain are constrained by entanglements
even in the interface region.
Rather different results are obtained due to this
entanglement effect.
Finally we summarize our results and present some directions for
further investigation in the last section.

\section{Model and simulation procedures}
\label{model_and_simulation_procedures}

\subsection{Physical model}
\label{physical_model}

Now we describe the overall picture of the chain penetration process.
Initially the chain
is assumed to be exposed to the air and hence to be spread out on the
substrate with a thickness of order $a$, the monomer size. 
This is a result of the high surface tension in this situation.
Therefore the initial conformation on the interface
is assumed to be a 2~dimensional random walk. Throughout
this work we will ignore the interactions of the connector
chain with itself.

In most of this work,
the connector chain is assumed to be chemically
compatible with the constituent polymers of the elastomer.
We can quantify this by defining an affinity parameter 
between the chain and the wall, $\chi$,  
which is the interaction energy per monomer
between the chain and the wall
relative to that between the chain and the gel, in units of~$k_B T$.
$\chi$ is assumed to be close to zero.
In this case, the equilibrium conformation is plume-like, having
dimensions of order  that of a free phantom chain. If there were
no entanglements it should reach an equilibrium conformation
in a Rouse time scale.
However the topological constraints imposed by the network prevent the
chain from moving freely into the gel.
Initially the chain is repelled from the surface and makes
``hernias'' or kink-like configurations that penetrate into the network.
For the case of a chain with both ends tethered,
this penetration by making hernias is the only mode of motion.
Depending on the thickness of the slab and the chemical
properties of the interface, the chain may penetrate into the
network deep enough to contribute to the adhesion of the two materials.
But for the singly tethered chain,
the free end moves into the gel, at a point of order $N^{1/2}a$ from
the tethered end, forming a plume. We will see that the plume
pulls out much of the slack from the rest of the chain 
because this is entropically favored.

The plume size
that initially forms depends strongly on the degree of
entanglements in the interface.
Even though we concern ourselves mostly with the low coverage regime in
which case the chains should move freely inside the interface, 
we discuss in~\secn{\ref{interfacial entanglement_effects}}  the effects
of entanglements in the interface. These should
be present at higher densities and can be caused by interactions
between grafted chains 

Let us suppose that a chain initially penetrates into
the network at a distance~\mbox{$R\ne 0$} from the anchor site.
Then we will see shortly that this 
conformation is in a metastable state with a very large relaxation 
time\cite{KPOTCBM:MolecularVelcro}.
After the time scale associated with reaching
this metastable plume state,
the relaxation time depends exponentially on the penetrated
length of the tethered chain. At times greater than this the
chain relaxes by slowly shifting its entry point into the
gel and eventually the total toughness is recovered.

\subsection{Computer simulation}
\label{computer_simulation}

The cross-linked network is modeled in this work by the
so called {\em cage model}.
The cage model was first introduced by 
Doi\cite{MD:CageModel}
and Evans and Edwards\cite{KEESFE:ComputerSimulation:1,%
SFEKEE:ComputerSimulation:2,KEESFE:ComputerSimulation:3}
to investigate numerically the properties of the reptating polymers
predicted by de~Gennes\cite{PGdG:ReptationPolymer}.
The chain sits on a lattice with lattice constant $a$.
In between lattice points is a cage.
The cage which resembles a cubical ``jungle-gym''
imposes topological constraints on the motion of polymers.
The  distance between adjacent parallel bars is chosen 
to be the entanglement length~$d_e$,
which is usually assumed to be an integral multiple of
the lattice constant~$a$.

In this simulation, we only consider the case where the Kuhn
length is the same as the entanglement length of the network, 
that is,~$a=d_e$.
In this case, if we consider both  the cross-link points of the cage
and the lattice sites the monomers can occupy, we obtain
a lattice that is body centered cubic (bcc),
which is bipartite.
The cross-link points occupy one sublattice
and the polymer moving around it occupies the other.
The cage is restricted to $z \ge d_s$,
where $d_s$ is the thickness of the interface region.
Due to the discreteness of the lattice, $d_s$ is somewhat ambiguous
and we use half-integer values for $d_s$ because the cage lies half
way between lattice points.
In most experiments in the low coverage regime,  the
network is frequently in contact with the surface and thus $d_s=1/2$ 
should correspond well to this situation. (Recall $d_e=a=1$).
The case with general half-integer $d_s$'s 
will be discussed briefly in the next section.

We tether one end of the $N$-mer
chain at the origin and place the rest randomly
on the surface, which is chosen to be
the $xy$-plane with $z=0$. 
(The other end is also tethered in
\secn{\ref{chains_tethered_at_both_ends}}.)
After this it
moves into the gel network by dynamics that are described below.
The chain is not allowed to penetrate into the wall~$z<0$.

We choose an arbitrary bead other than the grafted end(s), and 
attempt to move it according to the following rules.
\begin{description}
\item[move I]  If it is the free end, it is moved to
one of the six nearest neighbor points
(on the same sublattice).
Excluded volume effects are completely ignored in this work except for
the presence of the hard wall.

\item[move II] If its two nearest neighbor beads 
along the backbone are on the
same lattice point, then we move this middle point randomly
to one of the six
nearest neighbor points.

\item[move III] In most of the cases we study there
are no entanglements on the surface and so 
we must add one more move set.
If the chosen point and its two nearest neighbor beads lie in the
region~$z < d_s$ and they form a right angle, 
we flip the middle bead
through the diagonal made by the other two.
\end{description}

If the newly selected point (after the temporary move) 
does not violate
the constraints imposed by the wall we accept the move and update the
conformation of the polymer.  In the course of a simulation
this procedure is repeated many times with  beads that are
chosen randomly.

In systems where there is an interaction between 
the chain and the wall, the case of
non-vanishing~$\chi$,
the acceptance probability is 
\[ \min(1,e^{-\frac{\Delta E}{k_B T}}) \]
following the
usual Metropolis algorithm.
Here $\Delta E$ is the energy change after the attempted move, which is
$\pm\chi$ times $k_BT$ 
if the number of beads which are in contact with the
wall changes by one.

We define a Monte Carlo Step (MCS) as $N-1$ 
such attempts for the singly grafted chain and as $N-2$
for the doubly grafted one.
One MCS corresponds to the monomeric relaxation time~$\tau_1$.

As mentioned in the previous section, the time dependence of the
number of monomers in the penetrated part of the chain plays
an important role in the toughness recovery process.
So does the
distance to the penetration point, $R$, in discussing the long time
behavior. 
We measure these as well as other
quantities such as the average height $z_{\rm av}$ and
the radius of gyration in the vertical direction $z_{\rm g}$. These are
defined by the following equations.
\begin{eqnarray}
z_{\rm av}  &=& \frac{1}{N} \sum^N_{i=1} z_i  \\
z_{\rm g}^2  &=& \frac{1}{N} \sum^N_{i=1} (z_i - z_{\rm av})^2
\end{eqnarray}
where $z_i$ is the $z$-coordinate of the $i$-th monomer.

In the following sections, we will analyze the dynamics
of this system when $\chi$ is zero, 
negative ($-\ln(2)$)  and positive ($\ln(10)$).
In order to find out how far away these systems are 
from equilibrium, it is useful to have equilibrium
statistics.
These are easily obtained by using 
faster dynamics that do not have the
topological constraints imposed by the network.
To obtain these faster dynamics, we include the
right angle flip (move III) everywhere along the chain. These
moves can pass cage bars but give precisely the same
equilibrium statistics.
These dynamics have the relaxation time of the Rouse time ($\sim N^2$)
and therefore give accurate results for equilibrium
quantities.
The equilibrium values of several quantities are tabulated
for $N=20$, $50$, $100$ and $200$, 
and these will be used in the following 
sections.
See tables~\ref{zero_chi}, \ref{attractive_0.7}
and~\ref{repulsive_2.3}.

   \begin{table}[tbh]
\begin{center}
\begin{tabular}{|c|c|c|c|c|}
\hline
length ($N$) & 20 & 50 & 100 & 200 \\
\hline
primitive chain length ($L_{\rm eq}$) & 12.5 & 32.5 & 66.5 & 132.5 \\
primitive length of the penetrated part ($ g $)
& 11.5 & 31.5 & 65.0 & 131.5 \\
radial distance to the entrance point ($ R $)
& 1.1 & 1.1 & 1.1 & 1.1 \\
number of monomers on the trailing part ($s$) 
& 3.8 & 3.6 & 3.5 & 3.5 \\
height ($ z_{\rm av} $)
& 1.5 & 3.0 & 4.4 & 7 \\
radius of gyration ($R_{\rm g}$)
& 1.0 & 2.8 & 4.1 & 5.7 \\
vertical component of $R_{\rm g}$ ($ z_{\rm g} $)
& 1.0 & 1.6 & 2.2 & 3.3 \\
\hline
\end{tabular}
\end{center}
\caption{Equilibrium values of various quantities when there is no
interaction between the chain and the wall, i.e., $\chi = 0$. These
were obtained using faster dynamics as explained in the text.
Note that $a$ is set to 1.}
\label{zero_chi}
\end{table}

   \begin{table}[tbh]
\begin{center}
\begin{tabular}{|c|c|c|c|c|}
\hline
length ($N$) & 20 & 50 & 100 & 200 \\
\hline
primitive chain length ($L_{\rm eq}$) 
& 6.0 & 14.5 & 30.5 & 66.5 \\
primitive length of the penetrated part ($ g $)
& 3.1 & 10.6 & 26 & 62 \\
radial distance to the entrance point ($ R $)
& 3.1 & 4.2 & 4.5 & 4.5 \\
number of monomers on the trailing part ($s$) 
& 15.5 & 28.5 & 36.5 & 39 \\
height ($ z_{\rm av} $)
& 0.14 & 0.13 & 0.12 & 0.12 \\
radius of gyration ($R_{\rm g}$)
& 1.7 & 2.6 & 3.7 & 4.9 \\
vertical component of $R_{\rm g}$ ($ z_{\rm g} $)
& 0.28 & 0.32 & 0.34 & 0.34 \\
\hline
\end{tabular}
\end{center}
\caption{Attractive case $\chi = -\ln(2)$.
While $R_{\rm g}$ is increasing as a function of $N$, there is no
substantial increase in $ z_{\rm g} $, indicating the equilibrium
conformation is pancake-like. Even though its thickness, 
$ z_{\rm av} $, is small, 
the penetration is appreciable in terms of $g$.}
\label{attractive_0.7}
\end{table}

   \begin{table}[tbh]
\begin{center}
\begin{tabular}{|c|c|c|c|c|}
\hline
length ($N$) & 20 & 50 & 100 & 200 \\
\hline
primitive chain length ($L_{\rm eq}$) 
& 13.5 & 33.5 & 67.0 & 133.5 \\
primitive length of the penetrated part ($ g $)
& 13.5 & 33.5 & 67.0 & 133.5 \\
radial distance to the entrance point ($ R $)
& 0.06 & 0.06 & 0.07 & 0.07 \\
number of monomers on the trailing part ($s$) 
& 1.09 & 1.08 & 1.08 & 1.08 \\
height ($ z_{\rm av} $)
& 2.3 & 3.8 & 5.5 & 8.0 \\
radius of gyration ($R_{\rm g}$)
& 1.9 & 2.9 & 4.1 & 5.8 \\
vertical component of $R_{\rm g}$ ($ z_{\rm g} $)
& 1.1 & 1.7 & 2.4 & 3.4 \\
\hline
\end{tabular}
\end{center}
\caption{Repulsive case $\chi = \ln(10)$.
The values obtained here is very close to those for $\chi =0$ except 
$R$ and $s$. The overall shape is vertically expanded slightly
compared to the non-interacting case.
In all these three cases, $g+R=L_{\rm eq}$.}
\label{repulsive_2.3}
\end{table}

Among those tabulated quantities the {\em the primitive length\/} 
deserves some discussion.
First, it is straightforward to see that the length of chain
that has  penetrated the network is an ill defined quantity.
It has an ambiguity for phantom chains
when kinks are present at the penetration point.
So, instead, we use 
the primitive path\cite{MDSFE86:TheoryPolymer}, 
which is defined as the path of minimum length
between the two end points that can be obtained by
pulling all the slack from a chain without violating any
topological constraints imposed by entanglements. 
The {\em primitive path length}, or {\em primitive length},
has been used in studying viscoelastic properties of polymers in a
melt or a gel and 
is proportional to the total chain length in equilibrium.
This yields an unambiguous definition of penetration length in our case:
after finding the primitive path, 
we divide it into two parts at the penetration point.
The number of beads in
its penetrated part will be denoted by $g$. It will be extensively
used in analyzing  our simulations.

\section{Chains tethered at both ends}
\label{chains_tethered_at_both_ends}

\subsection{Statics}
\label{statics}

We first study the case of a chain tethered at both ends.
We create a chain as a 2-d random walk on the surface
and fix both ends at these initial positions. 

Even though there is no excluded
volume interaction between the chain and the
gel network, the topological constraint imposed by permanently
attaching both ends of the chain to the substrate effectively
generates a repulsive interaction between them.
When the interface thickness $d_s$ is bigger than a certain critical
value, the penetration of the chain into the gel will be highly
suppressed. 
This phenomenon is an example of 
{\em ground state dominance}\cite{PGdG79:ScalingConcepts},
and in this case is due to a
purely topological reason\cite{MECJMD:ConjecturesStatistics}. 
In other words, when the chain enters the network its entropy is
reduced due to the topology of the doubly grafted chain which
restrict allowed conformations to only
double back on themselves. If the
interface is sufficiently thick, it maximizes its entropy by
staying in the interface.
However if the interface slab is thin enough, it loses
more entropy staying in the interface then venturing
into the network, and so it will be more favorable
for the chain to penetrate into the network
by making hernias. 
In such a case the height of the chain can grow 
beyond~$d_s$ depending on the amount of slack in the chain,
which is the total chain length subtracted from
the distance between the tethered ends. 
In fact, 
cross-over behavior from an extended state to the collapsed one is
expected when the thickness of the interface $d_s$ is increased.

\subsection{Dynamics}
\label{dynamics}

\begin{figure}[tbh]
   \begin{center}
   \ 
   \psfig{file=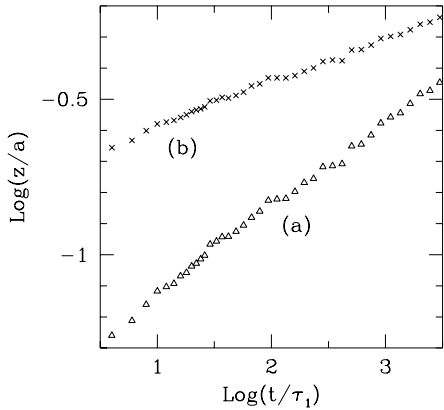,height=2.5in}
   \end{center}
\caption[Average height growth as function of time]{Log-log plot of
the initial expansion of the grafted chain with $N$=200.
Both ends of the chain were fixed at the surface and the data were
collected from the middle portion of the chain.
Here $z$ denotes either
(a) Average height $\langle z_{\rm av} \rangle$, or 
(b) Vertical component of the radius of
gyration $\langle z_{\rm g} \rangle$.
$\tau_1$ is defined to be 1 MCS.}
\label{AverageHeight}
\end{figure}

Upon attaching the rubber to the substrate,
the thickness of the disc-like conformation of the connector
increases to of order~$d_s$,
in a time scale of~$\tau_1 (\frac{d_s}{a})^4$%
\cite{PGdG79:ScalingConcepts}.
That is, 
the height $\langle z_{\rm av} \rangle$ averaged over the ensemble
grows as~$t^{1/4}$.
This is shown in \fig{\ref{AverageHeight}} for~$d_s = 1/2$.

Even though
this was obtained for a system in which both ends of the 
chain were fixed on the $xy$-plane,
this initial penetration behavior is expected to be the same for the
singly tethered case, because
the free end becomes important on a time scale of~$\tau_1 N^2$,
as we will see in the next section.
But in order to minimize boundary effects,
the data shown here was collected only from the middle 
~$N/2$ monomers.

\begin{figure}[tbh]
   \begin{center}
   \ 
   \psfig{file=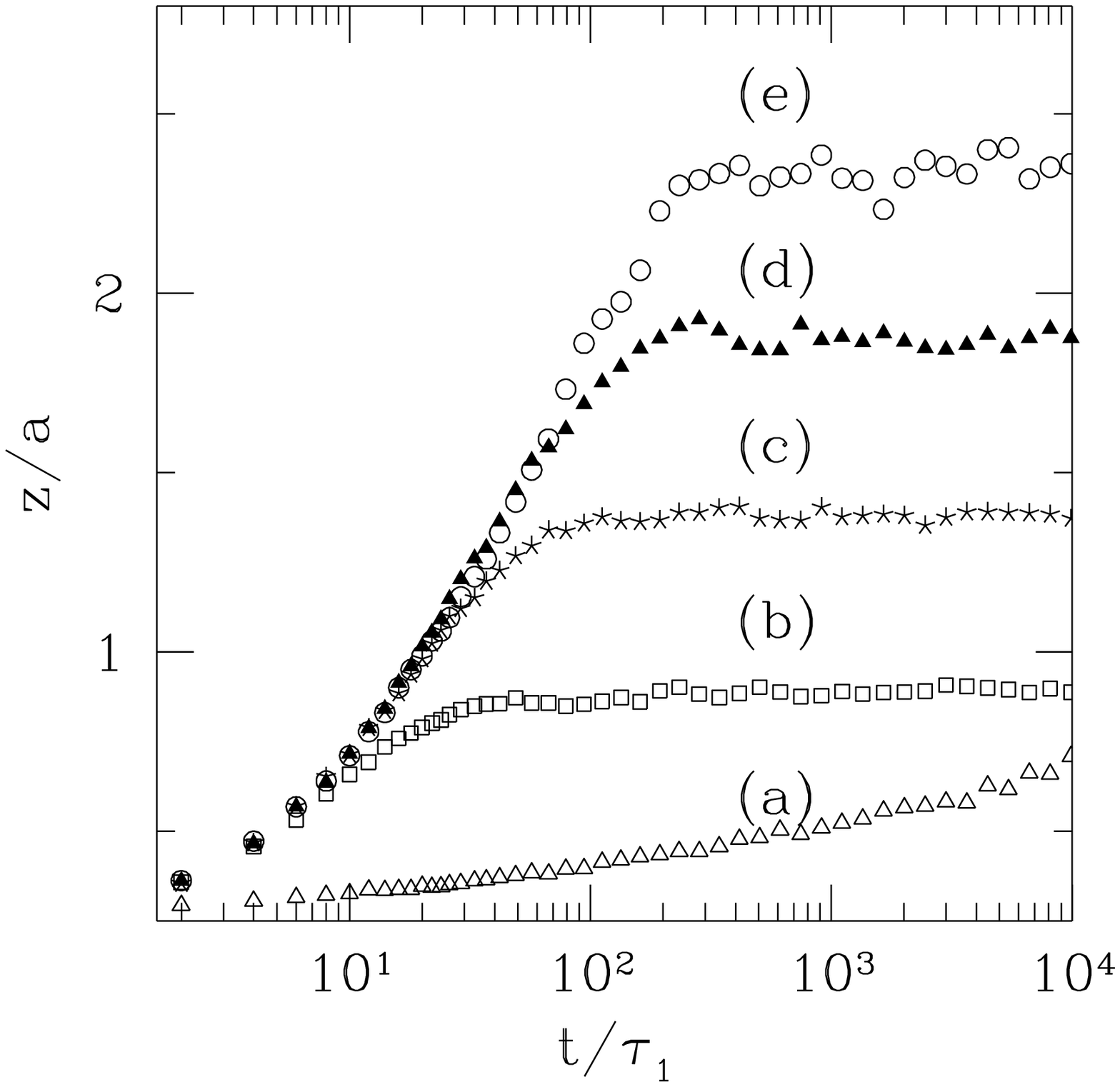,height=2.5in}
   \end{center}
\caption[Kink penetration as function of the slab
thickness]{Penetration of the hernias for various values of the slab
thickness $d_s$. \mbox{$N=100$}.
(a) $d_s=\frac{1}{2}$.
(b) $d_s=\frac{3}{2}$.
(c) $d_s=\frac{5}{2}$.
(d) $d_s=\frac{7}{2}$.
(e) $d_s=\frac{9}{2}$.
The vertical axis is
the average height~$\langle z_{\rm av} \rangle$ in units of $a$.
Without penetration $\langle z_{\rm av} \rangle$
is expected to be around
$d_s/2$ in equilibrium. Except for the case $d_s=\frac{1}{2}$,
there is no substantial penetration.}
\label{KinkPenetration}
\end{figure}

This initial rapid growth stops when the average height reaches 
of order~$d_s$. After this time scale the growth slows down.
\fig{\ref{KinkPenetration}}
shows the slow penetration of hernias into
the gel as a function of~$d_s$.
As stated earlier, penetration is achieved only for the
case of sufficiently small~$d_s$.
As the figure suggests, when $d_s$ is larger than~$1/2$
the saturation values of the {\em penetrated} height appear to 
be close to zero.

\begin{figure}[tbh]
   \begin{center}
   \ 
   \psfig{file=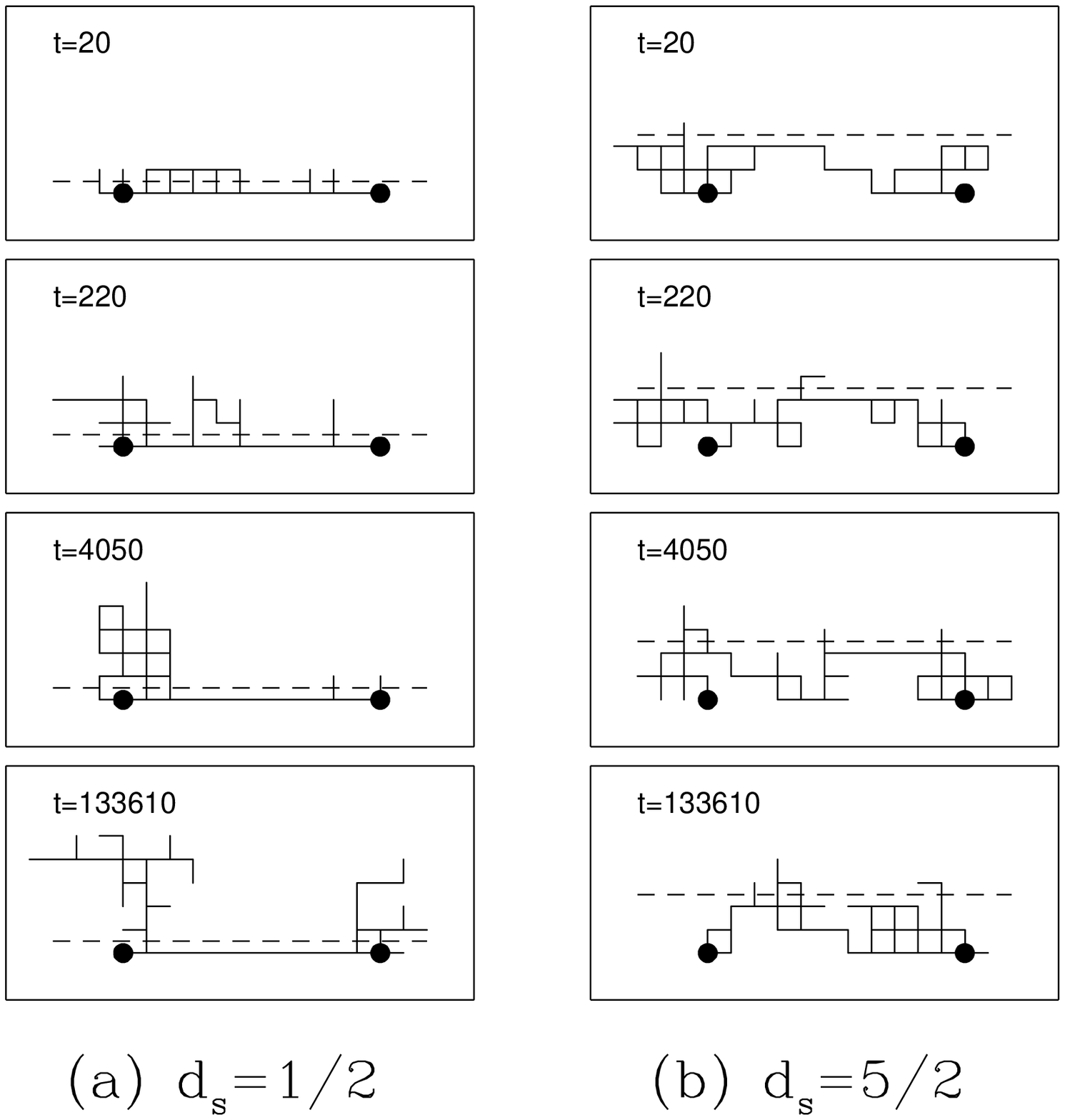,height=2.5in}
   \end{center}
\caption[Growth of hernias in 2-d system]{Relaxation of the 
doubly grafted
chain in 2 dimension. $N$=100.
(a) $d_s=1/2$
(b) $d_s=5/2$.
The two circles represent the tethered points and
the dotted lines represent thickness of the interface region.}
\label{GrowthHernias}
\end{figure}

\fig{\ref{GrowthHernias}} shows that this is indeed the case.
This picture was obtained from 2-d system 
for the sake of clarity.
There is a substantial
penetration for the case $d_s=\frac{1}{2}$, 
whereas for the case $d_s=2\frac{1}{2}$ 
there is very little penetration.

One interesting result is that the average
radius of gyration in the
$z$-direction, $\langle z_{\rm g} \rangle$,
has much weaker growth than the average height $\langle z_{\rm av} \rangle$ .
As shown in \fig{\ref{AverageHeight}}
the slope is $1/8$ rather than~$1/4$.
This shows that the chain as a whole translates away from the
wall, and penetrates the network at a much slower rate.
The $t^{1/8}$ increase in $\langle z_{\rm g} \rangle$
can be understood by noting that 
the loops in the network are lattice 
animals\cite{ARKSKN:PolymerChain}.
Lattice animals have a fractal dimension of 4 so that a loop in the
network of $m$ monomers has $\langle z_{\rm g} \rangle \propto m^{1\over 4}$.
The length of the loops increases by a diffusive process and therefore
$m \propto t^{1\over 2}$. These two equations lead to the exponent of
$1\over 8$ found above.

\section{Chains tethered at one end}
\label{chains_tethered_at_one_end}

\subsection{Theory}
\label{theory}

\subsubsection{Zero $\chi$}

In the previous section, it was shown that 
topological interactions limit the
penetration of the doubly tethered chain.
Topological
constraints do not limit conformations accessible to
the singly tethered case.
In this case the network, in our numerical model,
has no effect on equilibrium properties.
In addition, when
there is no attraction between the chain and the
surface, we expect the chain will eventually reach the
equilibrium plume-like shape mentioned earlier. As we now
describe, the network
affects the dynamics of this system in a drastic manner.
The effects described below were recently worked out by
O'Connor and McLeish\cite{KPOTCBM:MolecularVelcro}.

Once the chain end enters the gel,
that portion of the chain in the gel is confined to a tube
of diameter~$\sim\!d_e$.
The rest of the chain is constrained to the slab of 
thickness~$\sim\!d_s$.
Therefore we have two different relaxation modes, 
the snake-like motion of segments in the network
and the 2~dimensional Rouse motion for the rest.
The segment of length 
$g(t)a$ will assume a plume-like conformation
of size~$g(t)^\frac{1}{2} a$, as it enters the gel.
While this plume by definition starts at the surface,
the center of the plume diffuses
away from the wall because its size grows as more monomers
enter the gel.
Eventually the tension on the chain segment in the slab becomes big
enough to counteract the pulling force of the free end,
at which point the chain conformation will be 
``runner-like'', in that it consists of a plume and a trailing part
which connects the plume to the graft site.
This is a metastable state as claimed earlier.
Now we examine this process more quantitatively giving
a slightly simpler version of the argument of
O'Connor and McLeish\cite{KPOTCBM:MolecularVelcro}.

Let us suppose the free end penetrates into the gel at a distance $R$
from the origin. We will consider the statistical mechanics
of this situation when the penetration point is held fixed.
Then the free energy for the conformations, 
with $s$ monomers in the slab, can be written as follows,
\begin{equation}
\frac{F}{k_B T} = \alpha \frac{R^2}{s a^2} + \beta
	\frac{s a^2}{d_s^2}       \label{free_energy}
\end{equation}
The first term is the elastic energy of the portion of the
chain along the surface and the second term represents
the penalty due to its topological confinement in the 
slab\cite{PGdG79:ScalingConcepts}.
The free energy is defined relative to that of a plume in
true equilibrium, that is $R \sim d_s$.
The numerical coefficients 
$\alpha$ and $\beta$ are constants of order 1,
and will be retained because their values can have important
effects on the behavior of the system.
These constants depend only on
the geometrical properties of the system.
For the case with non-zero $\chi$, we have to add 
to \eqn{\ref{free_energy}}
one more term which is proportional to~$s$.
This is neglected for the moment.

By minimizing this expression with respect to $s$ we get
\begin{equation}
s = \left\{\begin{array}{ll}
	\frac{d_s R}{a^2}(\frac{\alpha}{\beta})^{\frac{1}{2}} &
	\mbox{if $R < \frac{Na^2}{d_s}
	(\frac{\beta}{\alpha})^\frac{1}{2}$} \\ 
	N & \mbox{otherwise} \end{array} \right.
\label{s_equil}
\end{equation}
Since there is no metastable state in the latter case,
the upper bound for $R$ is 
$\frac{Na^2}{d_s} (\frac{\beta}{\alpha})^{1/2}$.
When the chain does penetrate, 
the segment in the slab is highly stretched in this local minimum
because $s$ is linearly proportional to~$R$.
This implies that most of the slack in the
initial 2-d configuration of the chain
must flow out of the slab and end up in the penetrated
portion of the chain. This is a consequence of the unfavorable
entropic conditions at the surface.

This entropic driving force will then lead to
a rapid penetration of the chain by
the diffusion of kinks, or defects, along the backbone
of the chain. 
If we extend the concept of primitive path to include the 
slab\footnote{The primitive path in the slab is defined in this work
by a straight line between the two end points.}
in addition to the tube, one can see that the primitive length is
expected to increase in time. This is because the 
initial primitive length, $R \sim N^{1/2}$,
is shorter than the value after long times when 
the chain is in its equilibrium state, in which case the
primitive length, which is the sum of two parts one in the slab and
one in the gel, is proportional to $N$.
Hence the primitive length increases by diffusion of kinks.
The growth of the primitive length is then proportional to~$t^{1/2}$.

This process continues until
the spring force of the segment in the slab 
of length $sa = Na-ga$ becomes
comparable to the force due to confinement. 
The typical value of $s$ at which the growth of $g(t)$ slows down, 
is around $N^{1/2}$, i.e.,~$g \sim N$,
as will be shown shortly.
Hence $t^{1/2}$ growth persists until the time scale 
of~$\tau_1 N^2$.

We now would like to compute the free energy 
barrier $\Delta F$ that these
runner-like conformations must overcome in order to shift the
penetration point closer to the point of tether.
For these runner-like conformations, 
the free energy, as a function of the entrance point~$R$,
can be expressed as
\begin{equation}
\frac{F}{{k_B T}} = 2\frac{R}{d_s}(\alpha\beta)^{\frac{1}{2}}
\mbox{\hspace{0.4 in}}
(R < \frac{Na^2}{d_s}(\frac{\beta}{\alpha})^{1/2}) \label{free}
\end{equation}

The free energy barriers for these metastable states are
the free energy differences between the runner states and the
conformations when the whole chain is completely
pulled out from the gel, that is, when~$s=N$. Therefore
\begin{eqnarray}
\frac{\Delta F}{k_B T} &=& \alpha \frac{R^2}{N a^2} 
	+ \beta\frac{N a^2}{d_s^2} 
	- 2\frac{R}{d_s}(\alpha\beta)^{\frac{1}{2}} 
	\nonumber \\
	&=& \beta\left(\frac{a}{d_s}\right)^2 
	\frac{g^2}{N}. \label{delta_F}
\end{eqnarray}

The relaxation time of this metastable state is given by
Kramer's formula
\begin{eqnarray}
\tau_g &=& \tau_1 N^2 e^{\Delta F/k_B T}  \\
	&=& \tau_1 N^2 e^{\beta(\frac{a}{d_s})^2\frac{g^2}{N}}
\label{tau_g}
\end{eqnarray}
Compare this expression with the tube renewal time for a segment of
length $ga$ of the arm of a star polymer in a cross-linked 
network\cite{PGdG:ReptationStars,%
DSPEH:ViscoelasticProperties},
which is given by
\begin{equation}
\tau_{\rm arm} = \tau_1 N^2 e^{\beta'(\frac{a}{d_e})^2\frac{g^2}{N}}
\label{tau_arm}
\end{equation}
where the constant
$\beta'$ comes from the confinement of the chain in a tube of
diameter $d_e$ and will presumably be larger than~$\beta$%
\cite{MDSFE86:TheoryPolymer}.

Therefore we see that the plume formed at $R$ can tunnel 
into nearby local minima, in a time scale of~$\tau_g$,
thereby reducing the free energy given by \eqn{\ref{free}}.
This penetration process is finally completed when 
$R$ reaches of order~$d_s$.

\subsubsection{Non-zero $\chi$}
\label{analytical,non-zero chi}

In the previous section it was assumed that the connector chain 
was chemically
identical to the constituent polymers of the elastomer.
Now we discuss the more general situation of non-zero~$\chi$.

First, we study the attractive case, $\chi < 0$.
Attractive molecular forces,
such as van der Waals forces,
between the grafted chains and the
substrate will give rise to a negative $\chi$, as will
the chemical incompatibility between the chains and
the network.
The fact that the tethered chains have to swell the cross-linked network
is another reason for $\chi$ to become negative.
Because the surface represents an infinite potential barrier,
small negative $\chi$ is not enough to localize the chain to the
surface. There is a critical value of attraction $\chi^* <0$ 
giving two distinct
types of equilibrium behavior. For $\chi > \chi^*$ the chain behaves
much the same way
as described in the previous section when $\chi =0 $. 
For $\chi < \chi^*$ the chain will spread out on the surface 
in a pancake-like conformation,
giving rise to a finite
chain thickness $z_{\rm av}$ independent of chain length.
This is another situation characterized by ground state dominance.
%

Table~\ref{attractive_0.7} is obtained for~$\chi = - \ln(2)$.
The average height of the chain is about a quarter of~$d_s=a$
regardless of the size of the chain,~$N$,
indicating this $|\chi|$ is much bigger than $|\chi^*|$.
But even for such a strongly adsorbed case, 
the penetration is appreciable as shown in 
Table~\ref{attractive_0.7}, 
and we expect strongly adsorbed grafted chains to
make a significant contribution to the toughness of the interface.

Finally we mention the case with positive~$\chi$.
The case with $\chi=0$ is already a strongly repulsive situation,
since grafted chains are excluded from the wall 
because of the excluded volume constraint.
In fact the limit $\chi \rightarrow \infty$ is equivalent to
taking $\chi =0$ and making the wall thicker.
Therefore no qualitatively different results are expected in the
case of positive~$\chi$.
However the simulation results deviate from this expectation due
to a problem with Monte-Carlo dynamics. This will be
discussed in the next section.

\subsection{Simulation results}
\label{simulation_results}

\subsubsection{Zero $\chi$}

First we note that
for the bcc lattice cage model with $a=d_e$, analytic
solutions can be obtained exactly for many equilibrium 
properties\cite{EHDSP:StatisticsEntanglement}%
\cite{AMRJNetal:StatisticalTheory}.
Among others
the average primitive length in equilibrium, $L_{\rm eq}$, 
can be easily calculated using the fact that the primitive path is a
non-reversible random walk.
$L_{\rm eq}$ is given by
\begin{equation}
L_{\rm eq} = \frac{q-2}{q} N
\label{eq:exactl}
\end{equation}
where $q$ is the coordination number of the sublattice where the
polymers reside which is $q=6$ in our case, and so
\[
L_{\rm eq} = \frac{2}{3} N
\]
Simulation results given in Table~\ref{zero_chi}
are remarkably close to the above formula.

%
   
%

\begin{figure}[tbh]
   \begin{center}
   \ 
   \psfig{file=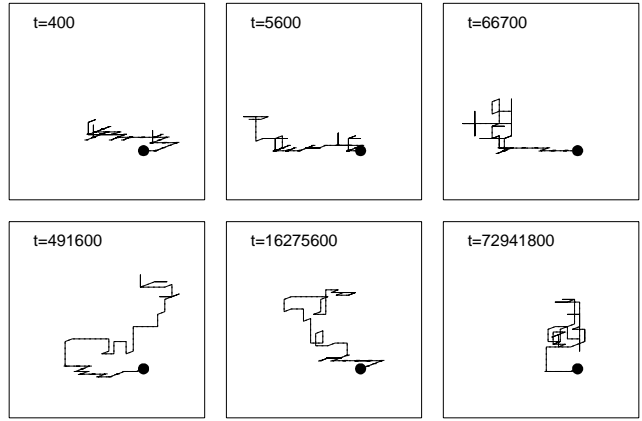,height=2.5in}
   \end{center}
\caption[Evolution of the tethered chain in a free
interface]{Penetration of the free end for $N$=50. $d_s=\frac{1}{2}$.
The tethered point is represented by a filled circle.
When $t=66700$ ($>\!\tau_1N^2$) we see the segment in the slab is
highly stretched. This is the runner-like conformation mentioned in
the text. The entry point into the rubber gets shortened as time goes
on.}
\label{EvolutionTethered}
\end{figure}

Now we discuss the penetration of the chain into the gel.
As was mentioned earlier,
this process can be understood as the diffusion of kinks out of the surface
region and into the plume region, and is driven by the entropy loss
in the vicinity of the surface.
The regime of rapid penetration of the chain is shown in 
\fig{\ref{EvolutionTethered}} for~$N=50$. 
Almost all the segments are inside
the gel by the time $t \sim 10^5 \tau_1$.

\begin{figure}[tbh]
   \begin{center}
   \ 
   \psfig{file=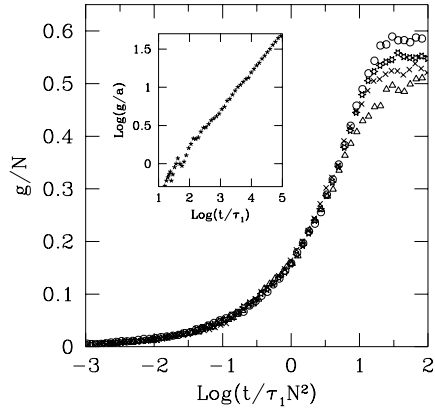,height=2.5in}
   \end{center}
\caption[Scaling of penetration length for different sizes of chains]{
Scaling plot for $N=20$, 50, 100 and 200. The vertical axis is 
the primitive length of the penetrated part, $g$,
scaled by
$N$ and the horizontal axis is logarithm of $t$
scaled by $\tau_1N^2$.
The inset is the log-log plot for $N$=200.
The exponent is 0.5.}
\label{ScalingPenetration}
\end{figure}

\fig{\ref{ScalingPenetration}}
demonstrates scaling behavior for chains of different
lengths, in this case $N=20$, 50, 100,
and~200. The vertical axis is scaled by the number of monomers, $N$,
and the time axis by~$N^2$, that is
\[
g(t) = N \tilde{g}(\frac{t}{\tau_1 N^2})
\]
Therefore the relaxation time is $\propto N^2$ as with
Rouse relaxation.  The log-log plot for $N=100$ is shown in the inset 
and fits well to a straight line with slope~$1/2$.

\begin{figure}[tbh]
   \begin{center}
   \ 
   \psfig{file=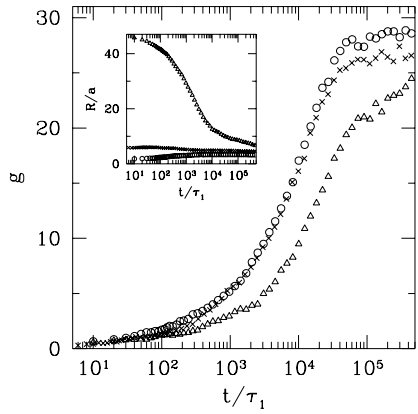,height=2.5in}
   \end{center}
\caption[Effect of initial conditions]{Comparison of the cases with
different initial conformations. $N=50$. $d_s = 1/2$.
The primitive length of the penetrated part, $g$, is shown for
compact (circle),
random walk (cross), and
stretched (triangle) initial conformations.
The inset shows the relaxation of the penetration point toward the
origin.  
After the Rouse time the relaxation is logarithmic 
for all three cases.}
\label{EffectInitial}
\end{figure}

This data was obtained, as stated earlier, with random walk initial
conformations, which probably differ somewhat from
experimental conditions.
However we have found that the result of $t^{1/2}$ growth 
is not very sensitive to the
initial conditions, because all but the typical
value of $R \sim N^{1/2} a$ are suppressed in the initial penetration
step.
If the chain is initially stretched, it contracts
first to a Gaussian shape to reduce the tension before the chain
enters the network.
As an extreme example demonstrating this point, 
we prepared ``samples'' in which the
grafted chains were initially stretched completely, say along the
$x$-axis.
The behavior for $g(t)$, was
very similar to the case with random walk initial conditions.
But, as we can see in \fig{\ref{EffectInitial}}, the initial
penetration was clearly suppressed compared to the previous case.
The inset compares the approaches of the
penetration points to the origin. 
In contrast, the samples with the chain initially bunched up near the
origin have a slight advantage. This head start is reflected in 
\fig{\ref{EffectInitial}} as a small left shift of the curve.

In a much longer time scale
after the chain reaches the first metastable state,
the relaxation is similar to that of a star polymer in a network.
In our case,
the penetration point relaxes slowly towards the origin by 
retracting the arm into the slab.

\begin{figure}[tbh]
   \begin{center}
   \ 
   \psfig{file=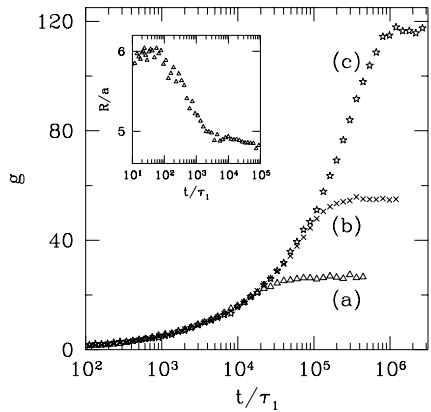,height=2.5in}
   \end{center}
\caption[Chain growth as a function of time]{Increase of the
penetration length as a function of time.
After the initial power-law growth, it crosses over to logarithmic
behavior.
(a) $N$=50.
(b) $N$=100.
(c) $N$=200.
The cross-over behavior is also apparent for the penetration point
relaxation shown in the inset for $N=50$.}
\label{ChainGrowth}
\end{figure}

Since the initial value of $R$ is typically $N^{1/2} a$,
the increase of $g$ at this last stage is
of order~$\frac{d_s}{a} N^{1/2}$,
which is negligible unless~$d_s \gg a$.
The system gains almost all its toughness in a 
time~$t \sim \tau_1 N^2$.
Note also that the increase of toughness after this time scale
proceeds very slowly because in situations with even moderate chain
length,
the characteristic time,
\(
\tau_1 N^2 e^{\beta(\frac{a}{d_s})^2N}
\)
is very large compared to~$\tau_1 N^2$.
\fig{\ref{ChainGrowth}}
shows that $g(t)$ increases until it reaches a
saturation value~$\sim\!N$.
The inset is for~$N=50$ and shows the penetration point approaching 
the origin in an exponential time scale. Systems with $N$ larger than
50 did not get to the last stage in a reasonable simulation time.

\subsubsection{Non-zero $\chi$}
\label{numerical,non-zero chi}

\begin{figure}[tbh]
   \begin{center}
   \ 
   \psfig{file=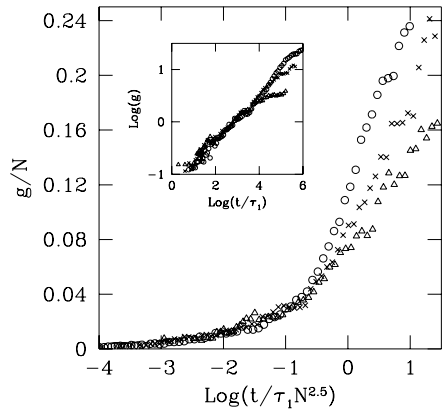,height=2.5in}
   \end{center}
\caption[Penetration of chain when the wall is attractive]{
Scaling plot for the attractive case, $\chi=-\ln(2)$.
$N$ is 20 (triangle), 50 (cross) and 100 (circle).
$g$ is scaled by $N$ and $t$ is scaled by $\tau_1 N^{2.5}$.
The inset shows the exponent is indeed close to 0.4.}
\label{PenetrationChain}
\end{figure}

\begin{figure}[tbh]
   \begin{center}
   \ 
   \psfig{file=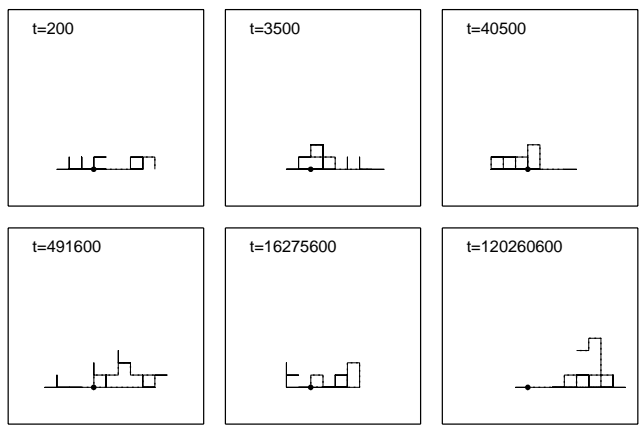,height=2.5in}
   \end{center}
\caption[Attractive case in 2 d]{
The typical relaxation modes for the attractive case with
$\chi = -\ln(2)$.
For clarity, the 2~dimensional picture is shown here. $N=50$.}
\label{AttractiveCase}
\end{figure}

First we examine the attractive case numerically. 
\fig{\ref{PenetrationChain}} shows the case 
$\chi = - \ln(2)$ for different chain lengths.
They scale as 
\[
g(t) = N \tilde{g}(\frac{t}{\tau_1 N^{5/2}})
\]
for short times, $t < \tau_1 N^{5/2}$,
indicating the growth is initially $t^{2/5}$.
The log-log plot in the inset shows
that the exponent is indeed close to 0.4.
The reason why the exponent for the attractive case is 0.4 
is unknown. It is also apparent
that the toughness increases rather quickly with time
again indicating that this situation should lead to an enhancement
of toughness in an experimentally accessible time scale.
This penetration process for the attractive case is shown in 
\fig{\ref{AttractiveCase}} for the 2~dimensional case.

When the substrate is another rubber-like material 
instead of an impenetrable surface, one has a similar situation.
Since the wall acts as a repulsive potential,
its removal decreases the effective value of~$\chi$, 
making the interaction attractive when the substrate is not
too chemically incompatible with connectors.
In this case, one has a large toughness
because the grafted chain prefers to stay near the interface
region while making ``many stitches'', of order~$N^{1/2}$,
between the two rubbers.
This situation was recently analyzed by 
Ji and de~Gennes\cite{HJPGdG:AdhesionConnector}.
%

Now we discuss the opposite extreme.
When the gel swelling is not very significant 
and the connector chain is compatible with the network, 
\(\chi\) is often positive.
In addition to the chemical disparity
between the chain and the substrate,
contamination of the
interface can be another reason for the chain to disfavor the surface.
In this case, the chain tends to be expelled from the surface more
strongly compared to the case~$\chi = 0$.
Paradoxically, our simulation results suggest that
this expulsion slows
down the dynamics of chain penetration, although we believe
this is an artifact of using Monte Carlo dynamics, as explained below.
This is similar to the problem of Monte Carlo dynamics for 
DNA gel electrophoresis in a strong field, which has been investigated
in recent years%
\cite{MOdlCJMDetal:ElectrophoresisPolymer,JMDJDR:SimulationHighly,%
JMDTLM:TheoreticalStudies}.

\begin{figure}[tbh]
   \begin{center}
   \ 
   \psfig{file=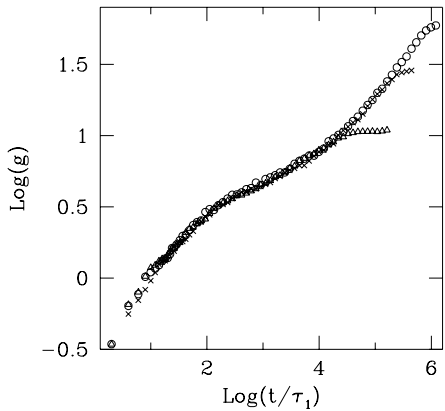,height=2.5in}
   \end{center}
\caption[When the wall is repulsive]{
Repulsive case with $\chi=\ln(10)$. 
$N$ is 20 (triangle), 50 (cross) and 100 (circle).
The initial growth is close to $t^{1/2}$. 
But the growth is slowed down gradually. 
After the Rouse time it regains momentum and $t^{1/2}$ growth resumes.
Finally it crosses over to logarithmic behavior.}
\label{WhenWall}
\end{figure}

Extruded defects driven by the strong surface repulsion grow until the tension
becomes large enough to counteract the repulsive force from the wall.
The chain at the wall, being under high tension, should be quite taut.
The lack of kinks at the wall means that the dynamics
become slow in this region. This is unphysical because  in reality even a
completely taut chain can diffuse, whereas the dynamics used
here do not have collective motions allowing a defect free segment to
move. Another way of seeing the difficulty is that a kink traveling
onto the surface must surmount an energy barrier of $\chi$ which
slows down the dynamics considerably when $\chi$ is large.
This is an artifact of Monte Carlo simulations with
local dynamics\cite{MOdlCJMDetal:ElectrophoresisPolymer}.
This difficulty can be overcome by either using
a Monte Carlo model with long range moves\cite{JMDJDR:SimulationHighly}
or using dynamics that are continuous in 
time\cite{JMDTLM:TheoreticalStudies}.
The situation here is not as serious as with electrophoresis where
potential energy scales linearly with chain length. The slowing
down here is a result of an activation free energy barrier $\propto
\chi$, with no dependence on the size of the defect.
Therefore on a sufficiently
long time scale the power law behavior should be the same as when
$\chi =0$ and a \(t^{1/2}\) growth is observed.
This interesting but spurious behavior is shown in \fig{\ref{WhenWall}}
for \(N = 20\), 50
and 100 with~\(\chi = \ln(10) \).

\section{Interfacial entanglement effects}
\label{interfacial entanglement_effects}

Now we study the effects of putting entanglements in
the interface region itself. 
There are some situations
where interfacial entanglements might become important. 
Suppose for example that
the substrate of the tethered chain embedded in a thin
layer containing a slightly incompatible
polymer network. Before adhesion to the compatible network,
the tethered chain is allowed to relax in the embedded layer. This
situation is similar to the case of negative $\chi$ discussed 
in \secn~\ref{analytical,non-zero chi}.
After some time the chain
will become intertwined with this layer of material.
Upon attachment of the compatible network it will move into it,
but now its motion is more highly constrained due to the
presence of entanglements in the thin layer.
We will now see that these surface entanglements have a
dramatic effect on the dynamics of the grafted chains.

The model studied in 
\secn{\ref{chains_tethered_at_one_end}}
must be modified to take into account
this entanglement effect in the interface region.
For simplicity we assume that the entanglement length
in this region is the same as in the network.
Also in the simulations, we are assuming as before, that
the lattice size of the cage equals the distance between monomers.
The tethered chain now moves along a small tube 
not only in the 3-d network 
but also in the 2-d interface region.

Before we consider the relaxation of this entangled chain, it should
be noted that the primitive length of the initial conformation is 
close to its equilibrium value, 
since it is already constrained to a tube. As one can
see from \eqn{\ref{eq:exactl}},  the 3-d value of the equilibrium
primitive length is $\frac{2}{3}N$, 
whereas the  2-d value is $\frac{1}{2}N$. 
This should be compared with the previous situation where the initial
primitive length was just the end to end distance of a Gaussian chain
in a slab which is of order $N^{1/2}$.

\begin{figure}[tbh]
   \begin{center}
   \ 
   \psfig{file=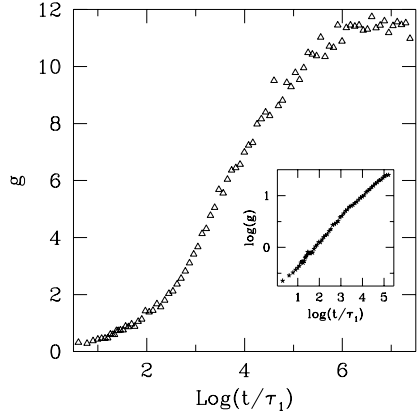,height=2.5in}
   \end{center}
\caption[Initial growth of self-entangled chains]{
Effect of entanglements in the slab. $d_s=0$.
The graph was obtained with
$N=20$. During the period $10^4 - 10^6$, the growth is logarithmic.
The inset with $N=100$
shows the initial growth has a $t^{1/2}$ dependence.}
\label{InitialGrowth}
\end{figure}

Because here the initial value of the primitive length is 25 percent
less than in equilibrium,
the chain slightly expands its primitive path while penetrating
into the gel.
Kinks on the surface feel a repulsive entropic force which causes
them to escape by climbing up the free end to avoid the surface.
The inset in 
\fig{\ref{InitialGrowth}} shows the initial growth of the plume. 
As the figure shows the exponent of $g$ as a function of
time is $\frac{1}{2}$, because this is a diffusive process. 
This regime lasts until \(t \sim \tau_1 N^2\).
By the end of this time 
the primitive length of the chain has increased by of order
$N$ monomers.

Viewed at this time scale, the end of the chain has formed a 
small plume and the
rest of the chain is in a tube embedded in the interface.
At longer time scales, the dynamics become very slow and similar to the
case of no interfacial entanglements.
In order for the tube to escape into the gel,
the chain has to overcome a free energy 
barrier as in the previous section.
From a relaxation time formula similar to
\eqn{\ref{tau_g}} or \eqn{\ref{tau_arm}}, we obtain
\begin{equation}
g(t) \approx \frac{d_e}{a}
	\left[N\ln(\frac{t}{\tau_1N^2})\right]^{\frac{1}{2}}
\end{equation}
This spans the major part of the healing process. In other words, only
a fraction of the toughness is recovered in a time scale $\tau_1 N^2$,
and to recover almost all the toughness takes a time exponential in
the chain length. This should be contrasted with the case of
no interfacial entanglements. There almost all the toughness is
recovered in the time scale $\tau_1 N^2$.

The entire relaxation behavior is shown in
\fig{\ref{InitialGrowth}} for~$N=20$.
A cross-over to logarithmic behavior after $t \sim 10^3 \tau_1$ is
evident in this figure.

\begin{figure}[tbh]
   \begin{center}
   \ 
   \psfig{file=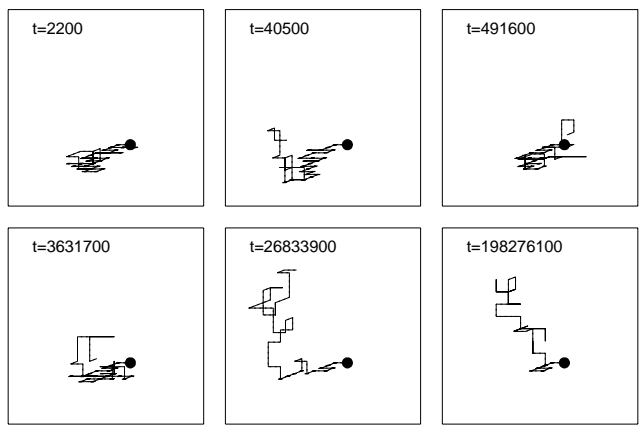,height=2.5in}
   \end{center}
\caption[Dynamics of the grafted chain near the interface]{The
vertical growth of the chain, when there is entanglement in the slab.
$N=50$.
The free end starts to penetrate slowly and the growth is logarithmic
after the time scale~$N^2$.
After $3 \times 10^6$ MCS's, 
appreciable amount of chain segments are still at
the interface.}
\label{DynamicsGrafted}
\end{figure}

\fig{\ref{DynamicsGrafted}} is to be compared with 
\fig{\ref{EvolutionTethered}}. 

\section{Conclusion}
\label{conclusion}

We have investigated the healing process of 
the interface between a cross linked elastomer and a grafted
solid.
As the grafted chain penetrates into the network the toughness
increases due to entanglements.
The dynamics of the grafted polymer was studied by Monte Carlo
simulation under various conditions.
In the case of $\chi \ge 0$ the
general features have been found to be in good agreement with the 
predictions of
O'Connor and McLeish\cite{KPOTCBM:MolecularVelcro}.

As soon as the contact is made, the grafted chain starts to penetrate
into the rubber network at a distance of order
$\sim\!N^{1/2}a$ from the anchor point.
If there is no interaction between the grafted chain and the surface,
the primitive length in the penetrated part increases as~$t^{1/2}$.
This process persists up to a time scale~$\tau_1 N^2$. In the
case of no interfacial entanglements, almost all the chain
segments penetrate into the network in this time scale.
After this rapid penetration stage, logarithmic relaxation follows.
When attractive interactions between the chain and
the substrate were added, the initial dynamics were found to be slower.
In the attractive case, the power scaling law is 
\mbox{$g(t)\sim t^{2/5}$}.

If we add topological constraints in the interface, the relaxation has
been found to be logarithmic
after the time scale~$\tau_1 N^2$. The dynamics become similar to
the arm of a star polymer, which has logarithmic relaxation. The
time to reach full toughness in this case
can easily exceed days or weeks in quite reasonable experimental
situations, using moderate size polymers. We can conclude that
toughness recovery is very sensitive to the conditions in the
interface. In order to get the best result for adhesion, we have to
eliminate the possibility of entanglements in the surface before 
attachment.

In this work, the wall did not play a crucial role.
One expects qualitatively similar behavior
for the adhesion between two different polymer networks.
Similar relaxation behavior is expected for a grafted chain. 
But the interpretation for the toughness recovery
should be modified because this is a many stitch problem as
discussed earlier.

The case of tethered polymers at interfaces under shear is more
difficult to simulate. The problem of large tensions
causes problems with Monte Carlo dynamics as mentioned earlier. 
These can be
overcome by using long range moves\cite{JMDJDR:SimulationHighly}, 
or continuous time dynamics\cite{JMDTLM:TheoreticalStudies}.  
We are currently investigating both of these techniques.

\section*{Acknowledgment}

We would like to thank H.R. Brown for suggesting this
problem and for several valuable discussions.
This work was supported by NSF Grant DMR-9112767.


\begin{thebibliography}{10}

\bibitem{ERPGdG:RubberRubber}
E.~Rapha{\"{e}}l and P.-G. de~Gennes.
\newblock Rubber-rubber adhesion with connector molecules.
\newblock {\em J.Phys.Chem.}, 96:4002--4007, 1992.

\bibitem{PGdG:WeakAdhesive}
P.-G. de~Gennes.
\newblock Weak adhesive junctions.
\newblock {\em J.Phys.France}, 50:2551--2562, 1989.

\bibitem{HRB:AdhesionPolymers}
H.R. Brown.
\newblock The adhesion between polymers.
\newblock {\em Annu.Rev.Mater.Sci.}, 21:463--489, 1991.

\bibitem{HJPGdG:AdhesionConnector}
Hong Ji and P.-G. de~Gennes.
\newblock Adhesion via connector molecules: The many-stitch problem.
\newblock {\em Macromolecules}, 26:520--525, 1993.

\bibitem{KEE:ScalingAnalysis}
K.E. Evans.
\newblock A scaling analysis of the fracture mechanisms in glassy polymers.
\newblock {\em J.Poly.Sci.Polym.Phys.}, 25:353--368, 1987.

\bibitem{KPOTCBM:MolecularVelcro}
K.P. O'Connor and T.C.B. McLeish.
\newblock Molecular velcro: Dynamics of a constrained chain into an elastomer
  network.
\newblock {\em Macromolecules}, 26:7322--7325, 1993.

\bibitem{WFRHRB:ReicherAndBrown}
W.F. Reichert and H.R. Brown.
\newblock {\em Polymer}, 34(11):2289, 1993.

\bibitem{CCHRBetal:MolecularWeight}
Costantino Creton, Hugh~R. Brown, and Kenneth~R. Shull.
\newblock Molecular weight effects in chain pull-out.
\newblock IBM preprint, 1993.

\bibitem{HRB:EffectsChain}
Hugh~R. Brown.
\newblock Effects of chain pull-out on adhesion of elastomer.
\newblock {\em Macromolecules}, 26(7):1666--1670, 1993.

\bibitem{KLJKKetal:JKRExperiment}
K.L. Johnson, K.~Kendall, and A.D. Roberts.
\newblock {\em Proc.R.Soc.London A}, 324:301, 1971.

\bibitem{HRBCYHetal:InterplayIntermolecular}
Hugh~R. Brown, Chung-Yuen Hui, and Elie Rapha{\"{e}}l.
\newblock Interplay between intermolecular interactions and chain pullout in
  the adhesion of elastomers.
\newblock {\em Macromolecules}, 27(2):608--609, 1994.

\bibitem{MD:CageModel}
M.~Doi.
\newblock {\em Polym.J.}, 5:288, 1973.

\bibitem{KEESFE:ComputerSimulation:1}
K.E. Evans and S.F. Edwards.
\newblock Computer simulation of the dynamics of highly entangled polymers:
  Part 1 equilibrium dynamics.
\newblock {\em J.Chem.Soc.Faraday Trans. II}, 77:1891--1912, 1981.

\bibitem{SFEKEE:ComputerSimulation:2}
S.F. Edwards and K.E. Evans.
\newblock Computer simulation of the dynamics of highly entangled polymers:
  Part 2 static properties of the primitive chain.
\newblock {\em J.Chem.Soc.Faraday Trans. II}, 77:1913--1927, 1981.

\bibitem{KEESFE:ComputerSimulation:3}
K.E. Evans and S.F. Edwards.
\newblock Computer simulation of the dynamics of highly entangled polymers:
  Part 3 dynamics of the primitive chain.
\newblock {\em J.Chem.Soc.Faraday Trans. II}, 77:1929--1938, 1981.

\bibitem{PGdG:ReptationPolymer}
P.-G. de~Gennes.
\newblock Reptation of a polymer chain in the presence of fixed obstacles.
\newblock {\em J.Chem.Phys.}, 55(2):572--579, 1971.

\bibitem{MDSFE86:TheoryPolymer}
M.~Doi and S.F. Edwards.
\newblock {\em The Theory of Polymer Dynamics}.
\newblock Clarendon Press, 1986.

\bibitem{PGdG79:ScalingConcepts}
Pierre-Gilles de~Gennes.
\newblock {\em Scaling Concepts in Polymer Physics}.
\newblock Cornell University Press, 1979.

\bibitem{MECJMD:ConjecturesStatistics}
M.E. Cates and J.M. Deutsch.
\newblock Conjectures on the statistics of ring polymers.
\newblock {\em J.Phys.France}, 47:2121--2128, 1986.

\bibitem{ARKSKN:PolymerChain}
A.R. Khokhlov and S.K. Nechaev.
\newblock Polymer chain in an array of obstacles.
\newblock {\em Physics Letters}, 112A(3,4):156--160, 1985.

\bibitem{PGdG:ReptationStars}
P.G. de~Gennes.
\newblock Reptation of stars.
\newblock {\em J.Phys.France}, 36(12):1199--1203, 1975.

\bibitem{DSPEH:ViscoelasticProperties}
Dale~S. Pearson and Eugene Helfand.
\newblock Viscoelastic properties of star-shaped polymers.
\newblock {\em Macromolecules}, 17:888--895, 1984.

\bibitem{EHDSP:StatisticsEntanglement}
Eugene Helfand and Dale~S. Pearson.
\newblock Statistics of the entanglement of polymers: Unentangled loops and
  primitive paths.
\newblock {\em J.Chem.Phys.}, 79:2054--2059, 1983.

\bibitem{AMRJNetal:StatisticalTheory}
Anita Mehta, R.J. Needs, and D.J. Thouless.
\newblock A statistical theory of entangled lattice polymers.
\newblock {\em Europhysics Letters}, 14(2):113--117, 1991.

\bibitem{MOdlCJMDetal:ElectrophoresisPolymer}
M.~Olvera de~la Cruz, J.M. Deutsch, and S.F. Edwards.
\newblock Electrophoresis of a polymer in a strong field.
\newblock {\em Phys.Rev.A}, 33:2047--2055, 1986.

\bibitem{JMDJDR:SimulationHighly}
J.M. Deutsch and J.D. Reger.
\newblock Simulation of highly stretched chains using long-range monte carlo.
\newblock {\em J.Chem.Phys.}, 95(3):2065--2071, 1991.

\bibitem{JMDTLM:TheoreticalStudies}
J.M. Deutsch and T.L. Madden.
\newblock Theoretical studies of {DNA} during gel electrophoresis.
\newblock {\em J.Chem.Phys.}, 90(4):2476--2485, 1988.

\end{thebibliography}
\end{document}